\documentclass[]{spie}  

\usepackage{amsmath,amsfonts,amssymb}
\usepackage{graphicx}
\usepackage[colorlinks=true, allcolors=blue]{hyperref}

\title{Calibration of the instrumental polarization effects of SCExAO-CHARIS’ spectropolarimetric mode\thanks{~~Based on data collected at Subaru Telescope, which is operated by the National Astronomical Observatory of Japan.}}

\author[a]{Rob G. van Holstein}
\author[a]{Steven P. Bos}
\author[a]{Jasper Ruigrok}
\author[b]{Julien Lozi}
\author[b,c,d,e]{Olivier Guyon}
\author[f]{Barnaby Norris}
\author[a]{Frans Snik}
\author[g]{Jeffrey Chilcote}
\author[b,h,i]{Thayne Currie}
\author[j]{Tyler D. Groff}
\author[a]{Joost 't Hart} 
\author[k]{Nemanja Jovanovic}
\author[l]{Jeremy Kasdin}
\author[b,e]{Tomoyuki Kudo}
\author[m]{Frantz Martinache}
\author[n]{Ben Mazin}
\author[b,o]{Ananya Sahoo}
\author[e,p,q]{Motohide Tamura}
\author[b,e,r]{S\'ebastien Vievard}
\author[n]{Alex Walter}
\author[q]{Jin Zhang}

\affil[a]{Leiden Observatory, Leiden University, PO Box 9513, 2300 RA Leiden, The Netherlands}
\affil[b]{Subaru Telescope, National Astronomical Observatory of Japan, National Institutes of
Natural Sciences (NINS), 650 North A'oh\=ok\=u Place, Hilo, HI, 96720, U.S.A.}
\affil[c]{Steward Observatory, University of Arizona, Tucson, AZ, 85721, U.S.A.}
\affil[d]{College of Optical Sciences, University of Arizona, Tucson, AZ 85721, U.S.A.}
\affil[e]{Astrobiology Center of NINS, 2-21-1, Osawa, Mitaka, Tokyo, 181-8588, Japan}
\affil[f]{Sydney Institute for Astronomy, Institute for Photonics and Optical Science, School of
Physics, University of Sydney, NSW 2006, Australia}
\affil[g]{Univeristy of Notre Dame, Notre Dame, IN 46556, U.S.A.}
\affil[h]{NASA-Ames Research Center, Moffett Blvd., Moffett Field, CA, USA}
\affil[i]{Eureka Scientific, 2452 Delmer Street Suite 100, Oakland, CA, USA}
\affil[j]{Goddard Space Flight Center, 8800 Greenbelt Rd, Greenbelt, MD 20771, U.S.A}
\affil[k]{California Institute of Technology, 1200 E California Blvd, Pasadena, CA 91125, U.S.A.}
\affil[l]{Princeton University, Princeton, NJ 08544, U.S.A.}
\affil[m]{Laboratoire Lagrange, Universit\'e C$\mathrm{\hat{o}}$te d'Azur, Observatoire de la C$\mathrm{\hat{o}}$te d'Azur, CNRS, Parc Valrose, B$\mathrm{\hat{a}}$t. H. FIZEAU, 06108 Nice, France}
\affil[n]{University of California Santa Barbara, Santa Barbara, CA 93106, U.S.A.}
\affil[o]{Sokendai, Graduate University for Advanced Studies, Kanagawa Prefecture, Miura District,
Hayama, Shonankokusaimura, 240-0193, Japan}
\affil[p]{National Astronomical Observatory of Japan, National Institutes of Natural Sciences (NINS),
2 Chome-21-1 Osawa, Mitaka, Tokyo 181-0015, Japan}
\affil[q]{University of Tokyo, 7 Chome-3-1 Hongo, Bunkyo City, Tokyo 113-8654, Japan}
\affil[r]{Observatoire de Paris, LESIA, 5 Place Jules Janssen, 92190 Meudon, France}

\authorinfo{Further author information: (Send correspondence to RGvH) \\ RGvH: E-mail: vanholstein@strw.leidenuniv.nl}

\begin{document} 
\maketitle

\begin{abstract}
SCExAO at the Subaru telescope is a visible and near-infrared high-contrast imaging instrument employing extreme adaptive optics and coronagraphy. 
The instrument feeds the near-infrared light (JHK) to the integral field spectrograph CHARIS. 
Recently, a Wollaston prism was added to CHARIS’ optical path, giving CHARIS a spectropolarimetric capability that is unique among high-contrast imaging instruments. 
We present a detailed Mueller matrix model describing the instrumental polarization effects of the complete optical path, thus the telescope and instrument.
The 22 wavelength bins of CHARIS provide a unique opportunity to investigate in detail the wavelength dependence of the instrumental polarization effects. 
From measurements with the internal light source, we find that the image derotator (K-mirror) produces strong wavelength-dependent crosstalk, in the worst case converting ${\sim}95\%$ of the incident linear polarization to circularly polarized light that cannot be measured. 
Theoretical calculations show that the magnitude of the instrumental polarization of the telescope varies with wavelength between approximately 0.5\% and 0.7\%, and that its angle is exactly equal to the altitude angle of the telescope.
We plan to more accurately determine the instrumental polarization of the telescope with observations of a polarization standard star, and fit more comprehensive physical models to all experimental data.
In addition, we plan to integrate the complete Mueller matrix model into the existing CHARIS post-processing pipeline, with the aim to achieve a polarimetric accuracy of ${<}0.1\%$ in the degree of linear polarization.
Our calibrations of CHARIS’ spectropolarimetric mode will enable unique quantitative polarimetric studies of circumstellar disks and planetary and brown dwarf companions. 
\end{abstract}


%
%

\section{Introduction}\label{sec:intro}

The near-infrared (NIR) polarimetric modes of the high-contrast imaging instruments SPHERE-IRDIS\cite{beuzit2019sphere, dohlen2008infra, de2020polarimetric, van2020polarimetric} at the Very Large Telescope, Gemini Planet Imager (GPI)\cite{macintosh2014first, perrin2015polarimetry} at the Gemini South Telescope, and HiCIAO\cite{hodapp2008hiciao} at the Subaru Telescope have been very successful at imaging circumstellar disks of various ages\cite{garufi2017three, avenhaus2018disks, esposito2020debris, hashimoto2011direct, muto2012discovery} using polarimetric differential imaging (PDI).
GPI and SPHERE-IRDIS have also been used to search for polarization from planetary and brown dwarf companions\cite{millar2015beta, jensen2016point, van2017combining, jensenclem_exopol2}, leading to the detection of polarization from the stellar companion CS~Cha~B\cite{ginski2018first, haffert2020cs}, and more recently, the first detection of polarization from a planetary mass companion\cite{vanholstein_exopol2}.
In 2017, the Subaru telescope lost its high-spatial resolution NIR imaging polarimetric capability when HiCIAO was decommissioned. 
The current high-contrast imager is the Subaru Coronagraphic Extreme Adaptive Optics (SCExAO) system\cite{jovanovic2015subaru, lozi2018scexao}.
While initially SCExAO had no NIR polarimetric capabilities, recently a spectropolarimetric mode\cite{lozi2019new}, which is unique among high-contrast imagers, was implemented for its Coronagraphic High Angular Resolution Imaging Spectrograph (CHARIS)\cite{groff2017first} subsystem.

SCExAO is located on the NIR Nasmyth platform of the Subaru telescope, behind the AO188 system\cite{minowa2010performance}, which provides an initial low-order wavefront correction.  
The extreme adaptive optics system of SCExAO consists of a pyramid wavefront sensor that operates at wavelengths in the range 600-900~nm\cite{lozi2019visible}, a deformable mirror with 45 actuators over the pupil, and the real-time control software Compute and Control for Adaptive Optics\cite{guyon2018compute}. 
SCExAO feeds light to several science instruments, among which is CHARIS that provides low-resolution spectra with a resolving power $R = 18$ over the JHK-bands with a field-of-view (FOV) of $2"{\times}2"$.
To enable the spectropolarimetric capability, a Wollaston prism, which splits the light into two orthogonal linear polarization states, has been placed directly upstream of CHARIS.
The existing half-wave plate (HWP) located upstream of AO188, that was originally implemented for HiCIAO, is used to modulate the incident polarization to be measured. 

The accuracy of the CHARIS spectropolarimetric measurements is currently limited by instrumental polarization effects of the telescope and instrument.
The dominant effects are instrumental polarization (IP) and polarimetric crosstalk, which both cause the measured polarization state to differ from the true polarization state incident on the telescope. 
IP is the introduction of polarization signals by the telescope or instrument, and can make an unpolarized target appear to be polarized.
It is caused by the diattenuation of an optical component, which is the difference in reflectance or transmission of the two orthogonal linear polarization states.
Crosstalk is the conversion of linear to circular polarization (and vice versa) by optical components, and can lower the polarimetric efficiency (the fraction of the incident linear polarization that is actually measured) and can cause an offset of the measured angle of linear polarization. 
Crosstalk is the result of retardance, which is the difference in acquired phase between two orthogonal linear polarization states when light reflects of a mirror or transmits through an optic.

To enable unique and highly accurate quantitative polarimetry of circumstellar disks and planetary and brown dwarf companions with the spectropolarimetric mode of CHARIS, we are developing a detailed and fully validated Mueller matrix model that describes the instrumental polarization effects of the telescope and instrument.
For this work we follow the methodology and definitions from Ref.~\citenum{van2020polarimetric} as used for the polarimetric calibration of SPHERE-IRDIS. 
We calibrate the Mueller matrix model with measurements from SCExAO's internal light source and observations of polarization standard stars.
For these calibrations, the 22 wavelength bins of CHARIS provide a unique opportunity to investigate in detail the wavelength dependence of the instrumental polarization effects.
After completing the calibrations, we will use the model for the post-processing of science observations to correct for the IP and crosstalk. 
In this manuscript, we present the first results of this project.

The outline of this manuscript is as follows.
In \autoref{sec:theory} we describe the optical path of SCExAO-CHARIS and explain the Mueller matrix model that describes the instrumental polarization effects of the telescope and instrument. 
Subsequently, in \autoref{sec:MeasurementsDataReduction}, we outline the calibration measurements we performed with the internal light source, the data reduction, and the fitting of the model parameters to the calibration data. 
In \autoref{sec:results}, we then present the results of fitting the model parameters of the HWP and derotator, and discuss theoretical computations of the instrumental polarization introduced by the telescope.
Finally, in \autoref{sec:DiscussionOutlook} we present conclusions and a brief outlook on future steps for this project.

\section{Mathematical description of complete optical system}
\label{sec:theory}

\subsection{Optical path of SCExAO-CHARIS}
\label{subsec:OpticalPath}

\begin{figure}[!htb]
\begin{center}
\begin{tabular}{c}
\includegraphics[width=0.9\textwidth]{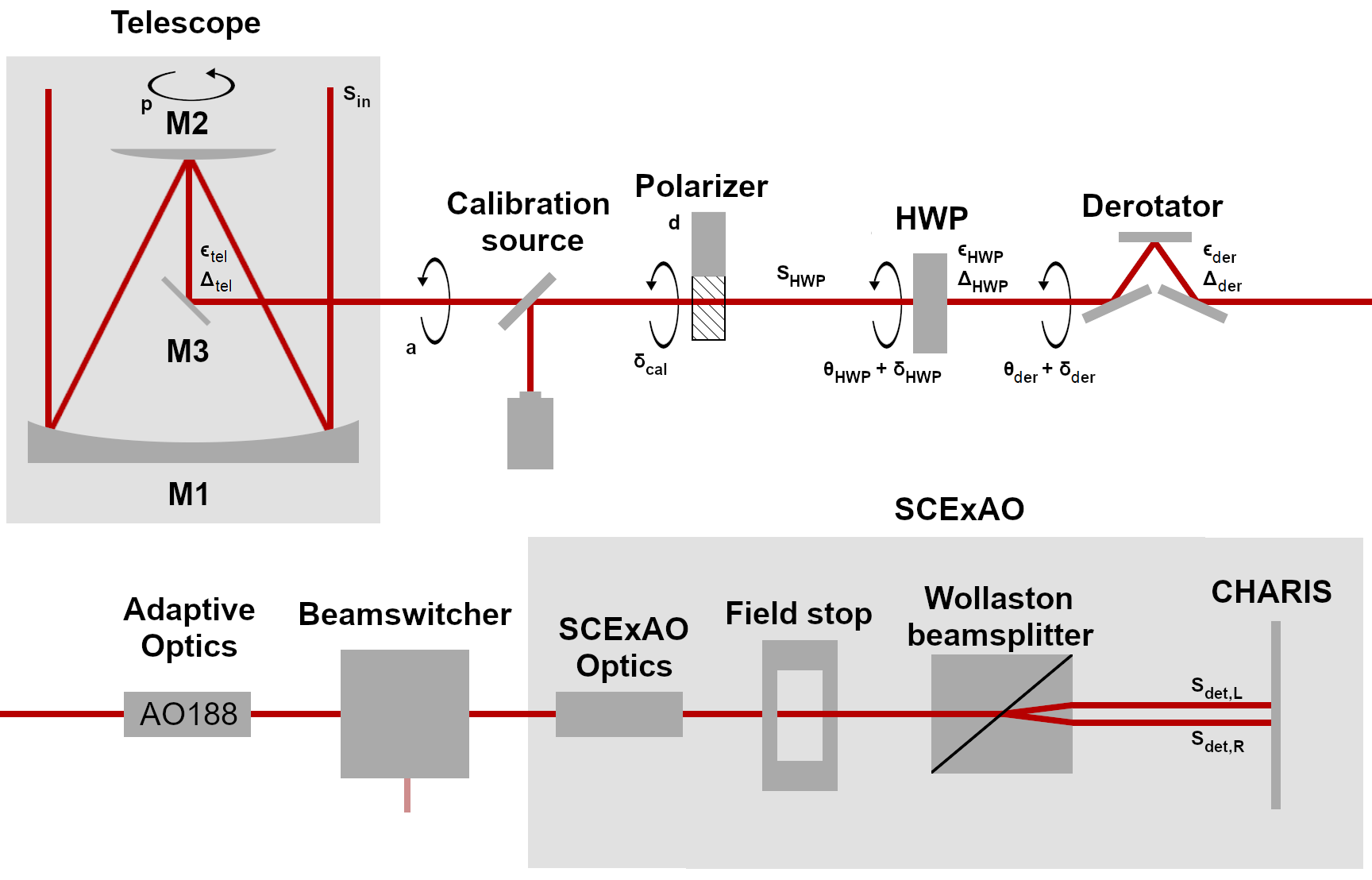}
\end{tabular}
\end{center}
\caption{
Overview of the optical path from the Subaru telescope to SCExAO-CHARIS, showing only the components relevant for polarimetric measurements. 
Circular arrows next to a component indicate that it is rotatable. 
The symbols next to the components indicate their model parameters and rotation (offset) angles.
Also shown are the Stokes vectors incident on the telescope, the HWP, and the detector.
Note that the beam switcher that is included in this figure has not been built at the moment of writing. } 
\label{fig:optical_path}
\end{figure}

A simplified optical path showing only the relevant components for the SCExAO-CHARIS polarimetric mode is presented in \autoref{fig:optical_path}.
The Subaru telescope is an 8-m class, altazimuth-mounted Ritchey-Chrétien telescope located at 4.2 km altitude on the summit of Mauna Kea. 
During observations the incident light is collected by the primary mirror (M1) and reflected to the secondary mirror (M2) that is suspended at the top of the telescope.  
The M2 subsequently reflects the light down toward the flat tertiary mirror (M3) that has an angle of incidence of 45$^{\circ}$ to reflect the light to the Nasmyth platform where SCExAO is located.
This M3 is coated with silver for high reflectivity in the NIR wavelength range.
While the object moves across the sky during the night, it rotates with the parallactic angle in the pupil of the telescope, and the telescope rotates with respect to the Nasmyth platform to track the altitude angle of the object.

When the light reaches the Nasmyth platform it passes a system that can be inserted into the light beam to perform calibration measurements. 
This calibration system consists of an internal (calibration) light source, a flat mirror with a 45$^{\circ}$ angle of incidence, and a manually rotatable linear polarizer.   
After the calibration system the light reaches an insertable and rotatable broadband half-wave plate (HWP), which is used to temporally modulate the polarization state of the incident light.
Following the HWP there is an image derotator, which consists of a three-mirror assembly (K-mirror) on a rotation stage.
When the K-mirror rotates, it rotates both the image and the angle of linear polarization.
For observations with SCExAO, the image derotator operates in pupil-stabilized mode. 
The derotator is followed by the adaptive optics of AO188, which includes a deformable mirror to correct for wavefront aberrations and a dichroic beamsplitter to feed the wavefront sensor.
All reflections in AO188 are in the horizontal plane (i.e., parallel to the Nasmyth platform).  

Currently, SCExAO is located directly behind AO188, and this location is shared by other instruments. 
Therefore, SCExAO has to be regularly craned in and out of this position for observations. 
As this process is very cumbersome, a beam switcher with multiple output ports to serve all the instruments on the Nasmyth platform is considered for a future upgrade\cite{lozi2017scexao}. 
SCExAO will then be located behind the beam switcher at one of these ports and the need for craning is eliminated.  
The reflections within the current beam switcher design are exclusively in the horizontal and vertical planes.

Within SCExAO the light has multiple reflections in the horizontal plane before a field stop limits the FOV to $2"{\times1}"$.
The field stop is followed by a Wollaston prism that splits the light into the orthogonal horizontal and vertical linear polarization states before it enters CHARIS.
Within CHARIS, the light passes a lenslet array and a prism to create a field of spectra in each of the two orthogonal linear polarization states, which then strike the detector side by side.

SCExAO hosts many different coronagraphs to suppress starlight, and most are not expected to affect polarimetric measurements. 
Two exceptions are the vector vortex coronagraph\cite{kuhn2018h} and the vector-Apodizing Phase Plate\cite{doelman2017patterned}.
However, with their current implementation these coronagraphs cannot be combined with polarimetry\cite{snik2014combining}.    

To measure the Stokes parameters $Q$ and $U$ and the corresponding total intensities $I_Q$ and $I_U$, a HWP cycle is performed which consists of four consecutive measurements with HWP switch angles equal to $0^\circ$, $45^\circ$, $22.5^\circ$, and $67.5^\circ$.
For each wavelength bin, we then compute from the resulting left (L) and right (R) intensities on the detector of each measurement, $I_\mathrm{det,L}$ and $I_\mathrm{det,R}$, the single difference and single sum as:
\begin{align} 
	X^\pm &= I_\mathrm{det,L} - I_\mathrm{det,R}, \label{eq:single_difference} \\[0.2cm]
	I_{X^\pm} &= I_\mathrm{det,L} + I_\mathrm{det,R}, \label{eq:single_sum} 
\end{align} 
with $X^\pm$ the single-difference $Q^+$, $Q^-$, $U^+$, and $U^-$ (taken at HWP switch angles $0^\circ$, $45^\circ$, $22.5^\circ$, and $67.5^\circ$), and $I_{X^\pm}$ the corresponding single-sum intensities $I_{Q^+}$, $I_{Q^-}$, $I_{U^+}$, and $I_{U^-}$.  
We then compute the double difference and double sum as:
\begin{align} 
	X = \dfrac{1}{2}\left(X^+ - X^-\right), \label{eq:double_difference} \\[0.2cm]
	I_X = \dfrac{1}{2}\left(I_{X^+} + I_{X^-}\right), \label{eq:double_sum}
\end{align} 
with $X$ the double-difference $Q$ or $U$ and $I_X$ the double-sum $I_Q$ or $I_U$.


\subsection{Mueller matrix model of optical path}
\label{subsec:MuellerMatrixModel}

To mathematically describe the instrumental polarization effects of the optical system of SCExAO-CHARIS, we create a Mueller matrix model similar to that of Ref.~\citenum{van2020polarimetric} for SPHERE-IRDIS.
We use the same definitions of the Stokes parameters, reference frames, and Mueller matrices as described in that work.
The relevant Stokes vectors and model parameters describing the components are shown in \autoref{fig:optical_path}.
Because the optical setups of SPHERE-IRDIS and SCExAO-CHARIS are quite similar, the Mueller matrix models of both instruments show many similarities.

The Stokes vectors reaching the left and right halves of the detector, $\boldsymbol{S}_\mathrm{det,L}$ and $\boldsymbol{S}_\mathrm{det,R}$, can be written in terms of the true Stokes vector incident on the telescope $\boldsymbol{S}_\mathrm{in}$ as:
\begin{align}
\begin{aligned} 
	\boldsymbol{S}_\mathrm{det,L/R} &= M_\mathrm{sys,L/R} \boldsymbol{S}_\mathrm{in}, \\[0.2cm]
	\begin{bmatrix} I_\mathrm{det,L/R} \\ Q_\mathrm{det,L/R} \\ U_\mathrm{det,L/R} \\ V_\mathrm{det,L/R} \end{bmatrix} &= \begin{bmatrix} I \hspace{-2pt}\rightarrow\hspace{-2pt} I & Q \hspace{-2pt}\rightarrow\hspace{-2pt} I & U \hspace{-2pt}\rightarrow\hspace{-2pt} I & V \hspace{-2pt}\rightarrow\hspace{-2pt} I~~\\~I \hspace{-2pt}\rightarrow\hspace{-2pt} Q & Q \hspace{-2pt}\rightarrow\hspace{-2pt} Q & U \hspace{-2pt}\rightarrow\hspace{-2pt} Q & V \hspace{-2pt}\rightarrow\hspace{-2pt} Q~~\\~I \hspace{-2pt}\rightarrow\hspace{-2pt} U & Q \hspace{-2pt}\rightarrow\hspace{-2pt} U & U \hspace{-2pt}\rightarrow\hspace{-2pt} U & V \hspace{-2pt}\rightarrow\hspace{-2pt} U~~\\~I \hspace{-2pt}\rightarrow\hspace{-2pt} V & Q \hspace{-2pt}\rightarrow\hspace{-2pt} V & U \hspace{-2pt}\rightarrow\hspace{-2pt} V & V \hspace{-2pt}\rightarrow\hspace{-2pt} V~~\end{bmatrix} \begin{bmatrix} I_\mathrm{in} \\ Q_\mathrm{in} \\ U_\mathrm{in} \\ V_\mathrm{in} \end{bmatrix},
\label{eq:mueller_matrix_definition}
\end{aligned}
\end{align} 
where $M_\mathrm{sys,L/R}$ is the Mueller matrix that describes the instrumental polarization effects of the images on the left or right half of the detector.
We can further express $M_\mathrm{sys,L/R}$ in terms of the separate components and their rotations as:
\begin{align}
\boldsymbol{S}_\mathrm{det,L/R} =& ~M_\mathrm{sys,L/R} \boldsymbol{S}_\mathrm{in}, \nonumber \\[0.2cm] 
	\boldsymbol{S}_\mathrm{det,L/R} =& ~M_\mathrm{W,L/R} T(-\varTheta_\mathrm{der}) M_\mathrm{der} T(\varTheta_\mathrm{der}) T(-\varTheta_\mathrm{HWP}) M_\mathrm{HWP} T(\varTheta_\mathrm{HWP}) \nonumber \\
	&~T(-a) M_\mathrm{tel} T(p) \boldsymbol{S}_\mathrm{in},
\label{eq:complete_model} 
\end{align} 
where $p$ is the parallactic angle, $a$ is the telescope altitude angle, and:
\begin{align}
\varTheta_\mathrm{HWP} &= \theta_\mathrm{HWP} + \delta_\mathrm{HWP}, \label{eq:theta_hwp}\\[0.2cm]
\varTheta_\mathrm{der} &= \theta_\mathrm{der} + \delta_\mathrm{der}, \label{eq:theta_der}
\end{align} 
with $\theta_\mathrm{HWP}$ the HWP angle, $\theta_\mathrm{der}$ the derotator angle, $\delta_\mathrm{HWP}$ the HWP offset angle, and $\delta_\mathrm{der}$ the derotator offset angle.
The rotations of the components and Stokes vectors are described with the rotation matrix $T$($\theta$):
\begin{equation} 
	T(\theta) = \begin{bmatrix} 1 & 0 & 0 & 0 \\ 0 & \cos(2\theta) & \sin(2\theta) & 0 \\ 0 & -\sin(2\theta) & \cos(2\theta) & 0 \\ 0 & 0 & 0 & 1 \end{bmatrix}.
	\label{eq:rotation_mueller_matrix} 
\end{equation}
The Mueller matrices $M_\mathrm{tel}$, $M_\mathrm{HWP}$, and $M_\mathrm{der}$ describe the three mirrors of the telescope, the HWP, and the three mirrors of the image derotator, respectively, and are defined as the component Mueller matrix $M_\mathrm{com}$:
\begin{equation} 
 	M_\mathrm{com} = \begin{bmatrix} ~1~ & ~\epsilon~ & 0 & 0 \\ ~\epsilon~ & ~1~ & 0 & 0 \\ ~0~ & ~0~ & \phantom{-}\sqrt{1 - \epsilon^2}\cos\varDelta & \sqrt{1 - \epsilon^2}\sin\varDelta \\ ~0~ & ~0~ & -\sqrt{1 - \epsilon^2}\sin\varDelta & \sqrt{1 - \epsilon^2}\cos\varDelta \end{bmatrix},
 	\label{eq:component_mueller_matrix} 
\end{equation}    
where $\epsilon$ is the diattenuation and $\varDelta$ is the retardance of the component.
The optical components downstream of the derotator, except for the Wollaston prism, to first order do not affect our measurements (see Ref~\citenum{van2020polarimetric}).
All reflections downstream of the derotator lie in the horizontal (and vertical) plane, and therefore these components do not produce crosstalk affecting the horizontal or linear polarization states that pass the Wollaston prism.
In addition, because we compute the Stokes parameters $Q$ and $U$ from the double difference, any diattenuation produced downstream of the derotator is removed.
The matrix $M_\mathrm{W,L/R}$ therefore only describes the left and right channels of the Wollaston prism:
\begin{equation} 
	M_\mathrm{W,L/R} = \dfrac{1}{2} \begin{bmatrix} 1 & \pm 1 & 0 & 0 \\ \pm 1 & 1 & 0 & 0 \\ 0 & 0 & 0 & 0 \\ 0 & 0 & 0 & 0 \end{bmatrix},
\label{eq:wollaston}	
\end{equation} 	
where the plus- and minus-signs are used for the left and right detector halves, respectively. 
This Mueller matrix represents a perfect polarizing beamsplitter, because the extinction ratio of Wollaston prisms exceeds 100.000:1 \cite{king1971some}.

To model the measurement of a Stokes parameter and the corresponding total intensity from a pair of measurements, we calculate for each measurement $\boldsymbol{S}_\mathrm{det,L}$ and $\boldsymbol{S}_\mathrm{det,R}$ from \autoref{eq:complete_model} and retrieve $I_\mathrm{det,L}$ and $I_\mathrm{det,R}$ from the first element of these vectors.
Subsequently, we compute the single differences $X^\pm$ and single sums $I_{X^\pm}$ from \autoref{eq:single_difference} and \autoref{eq:single_sum}, and the double-difference $X$ and double-sum $I_X$ from \autoref{eq:double_difference} and \autoref{eq:double_sum}.
Finally, we compute the normalized Stokes parameter $x$, as:
\begin{equation} 
	x = \dfrac{X}{I_X}.
\label{eq:normalized_stokes_parameter} 
\end{equation} 
We note that only for measurement pairs using HWP switch angles equal to $0^\circ$, $45^\circ$, $22.5^\circ$, and $67.5^\circ$ the Stokes parameters $X$ and $x$ correspond to $Q$, $U$, $q$, and $u$.

%
%

\section{Measurements and data reduction}
\label{sec:MeasurementsDataReduction}

With the Mueller matrix model of the optical system defined, we can now experimentally determine the model parameters describing the optical path downstream of the telescope.
To this end, we took a total of 340 calibration measurements with the internal (calibration) light source on February 19, 2020. 
The calibration polarizer, which is located downstream of the light source (see \autoref{fig:optical_path}), was inserted in the optical path with its transmission axis at $45^\circ$ to the horizontal.
This setup therefore injected almost 100\% positive $U$-polarized light into the system.
The measurements were taken at many combinations of HWP and derotator angles, with the HWP angle ranging from $0^\circ$ to $78.75^\circ$ with steps of $11.25^\circ$, and the derotator angle ranging from $45^\circ$ to $127.5^\circ$ in steps of $7.5^\circ$.
The aim of these measurements is to fit the retardances and offset angles of the HWP and the derotator.
We cannot use these measurements to fit the diattenuations of the HWP and the derotator, because for that we would need measurements without the calibration polarizer inserted.
However, because these diattenuations are expected to be very small (as is the case for SPHERE-IRDIS\cite{van2020polarimetric}) we did not take such measurements.

\begin{figure}[!htb]
\begin{center}
\begin{tabular}{c}
\includegraphics[width=0.7\textwidth]{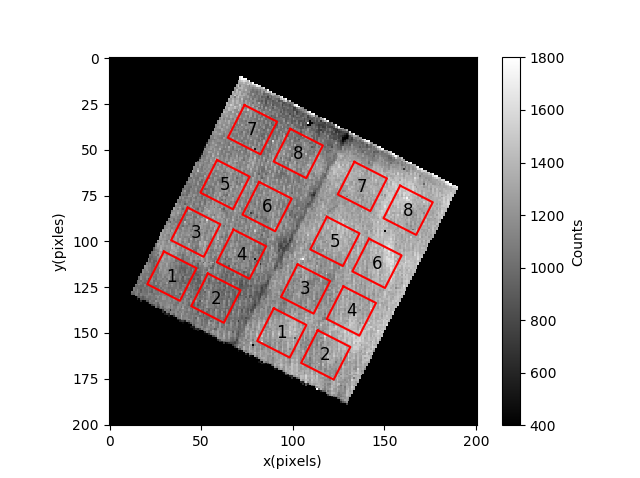}
\end{tabular}
\end{center}
\caption{Preprocessed image of a single wavelength bin showing the left and right $2"{\times}1"$ rectangular images of the two orthogonal linear polarization states. Eight square apertures, which are used to compute the single-difference and single-sum values, are shown in red on each rectangular image.} 
\label{fig:apertures}
\end{figure}

We preprocess the raw data with the CHARIS data reduction pipeline\cite{brandt2017data}$^{,}$\footnote{\url{http://princetonuniversity.github.io/charis-dep/}} using the standard settings (given by the examples in the documentation) for the wavelength calibration and data extraction.
The result is a three-dimensional data cube with two spatial axes and one spectral axis with 22 wavelength bins.   
One of the images resulting from the preprocessing (for a single wavelength bin) is shown in \autoref{fig:apertures}.
The square $2"{\times}2"$ image is divided into a left and right $2"{\times}1"$ rectangular image, corresponding to the two orthogonal linear polarization states. 

After the preprocessing is completed, we retrieve the normalized Stokes parameters from the images for each of the 22 wavelength bins.
To this end, we define eight square apertures on both the left and right halves of each image (see \autoref{fig:apertures}), and sum the flux in each of the apertures.
We then compute eight values for the single difference and single sum (following \autoref{eq:single_difference} and \autoref{eq:single_sum}) by respectively subtracting and adding the summed flux in apertures marked by the same number in \autoref{fig:apertures}. 
Subsequently, we compute values for the double difference and double sum (following \autoref{eq:double_difference} and \autoref{eq:double_sum}) from the single-difference and single-sum values computed from images that were taken at the same derotator angle and at HWP angles that differ by $45^\circ$.
Finally, we compute the normalized Stokes parameters by dividing the resulting double-difference values by the corresponding double-sum values (following \autoref{eq:normalized_stokes_parameter}).

To describe the measurements, we apply \autoref{eq:normalized_stokes_parameter} and insert \autoref{eq:single_difference} to \autoref{eq:wollaston}.
We only use the part of \autoref{eq:complete_model} downstream of the telescope:
\begin{align}
	\boldsymbol{S}_\mathrm{det,L/R} =& ~M_\mathrm{W,L/R} T(-\varTheta_\mathrm{der}) M_\mathrm{der} T(\varTheta_\mathrm{der}) T(-\varTheta_\mathrm{HWP}) M_\mathrm{HWP} T(\varTheta_\mathrm{HWP}) \boldsymbol{S}_\mathrm{HWP},
\label{eq:internal_model} 
\end{align} 
where $\boldsymbol{S}_\mathrm{HWP} = T(45^\circ - \delta_\mathrm{cal})[1, d, 0, 0]^T$ is the Stokes vector incident on the HWP, with $\delta_\mathrm{cal}$ and $d$ respectively the offset angle and diattenuation of the calibration polarizer. 

We now fit our model to the data points using non-linear least squares.
For this we use the Powell algorithm as implemented in the Python function \texttt{scipy.optimize.minimize}.
For each of the 22 wavelength bins we fit the retardance of the HWP $\varDelta_\mathrm{HWP}$, the retardance of the derotator $\varDelta_\mathrm{der}$, the HWP angle offset $\delta_\mathrm{HWP}$, and the diattenuation of the calibration polarizer $d$.
For the angle offsets of the derotator and the calibration polarizer, $\delta_\mathrm{der}$ and $\delta_\mathrm{cal}$, we fit only a single value valid for all wavelength bins. 
Because we cannot accurately determine the diattenuations of the derotator and HWP from these measurements with the calibration polarizer inserted, we set $\epsilon_\mathrm{HWP}$ and $\epsilon_\mathrm{der}$ to their ideal values of 0.

%
%

\section{Results}\label{sec:results}

\subsection{Instrumental polarization effects of the HWP and derotator}

\begin{figure}[!htb]
\begin{center}
\begin{tabular}{c}
\includegraphics[width=1.\textwidth]{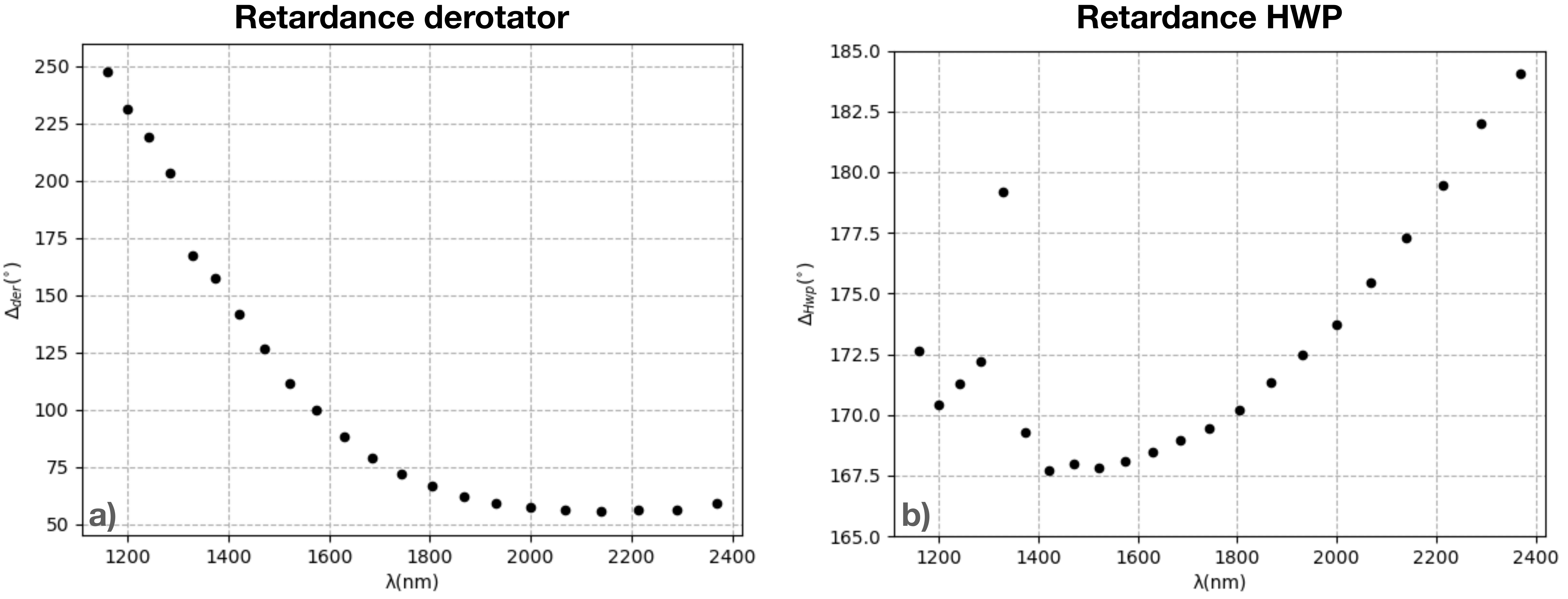}
\end{tabular}
\end{center}
\caption{
Fitted retardances of the a) derotator and b) the half-wave plate (HWP) as a function of wavelength.
Ideally, the derotator and HWP would have a retardance of 180$^{\circ}$ for all wavelengths.}
\label{fig:MeasuredRetardances}
\end{figure}

The fitted values for the retardance of the derotator and the HWP as a function of wavelength are shown in \autoref{fig:MeasuredRetardances}. 
The retardance of the derotator $\varDelta_\mathrm{der}$ would ideally have a constant value of 180$^{\circ}$. 
However, as shown in \autoref{fig:MeasuredRetardances}a, the retardance strongly varies with wavelength. 
At $\lambda = 1180$ nm the retardance has a value of $\varDelta_\mathrm{der} \approx 250^\circ$, then rapidly drops off to $\varDelta_\mathrm{der} \approx 70^\circ$ for $\lambda = 1800$ nm, and stabilizes around values between $60^\circ$ and $65^\circ$ for wavelengths between 1800 and 2400 nm.
Around $\lambda = 1600$ nm, the retardance is close to 90$^{\circ}$, which means that the derotator acts as an almost perfect quarter-wave plate (QWP) and therefore converts almost all incident linear polarization to circular polarization.
For the derotator offset angle, we assumed a constant value with wavelength and find $\delta_\mathrm{der} = -0.42^\circ$.

The HWP retardance $\varDelta_\mathrm{HWP}$ is plotted in \autoref{fig:MeasuredRetardances}b.
As with the derotator, the HWP retardance would ideally have a constant value of 180$^{\circ}$. 
However, we find that $\varDelta_\mathrm{HWP}$ varies between 167.5$^{\circ}$ and 185$^\circ$ for the measured wavelength range, and that it is exactly half wave around $\lambda = 2200$ nm.
The measured retardance curve is very similar to that of commercially available achromatic HWPs\footnote{\url{https://www.thorlabs.com/newgrouppage9.cfm?objectgroup_id=2193}}.
For $\lambda < 1400$ nm, the fitted retardance values are very noisy and do not follow a smooth curve as we would expect from the physics of wave plates. 
We have not yet determined the exact cause for this behavior. 
For our fit of the model parameters, we have allowed the HWP offset angle $\delta_\mathrm{HWP}$ to vary with wavelength and find that it indeed varies between $-0.4^\circ$ and $-0.8^\circ$. 
However, the orientation of the optic axis of an achromatic HWP cannot vary with wavelength due to the relatively simple design of such a wave plate. 
Therefore the wavelength variations of the HWP offset angle are most likely not physical and we adopt the mean value of $-0.6^\circ$ as the HWP offset angle for all wavelength bins.

\begin{figure}[!htb]
\begin{center}
\begin{tabular}{c}
\includegraphics[width=.5\textwidth]{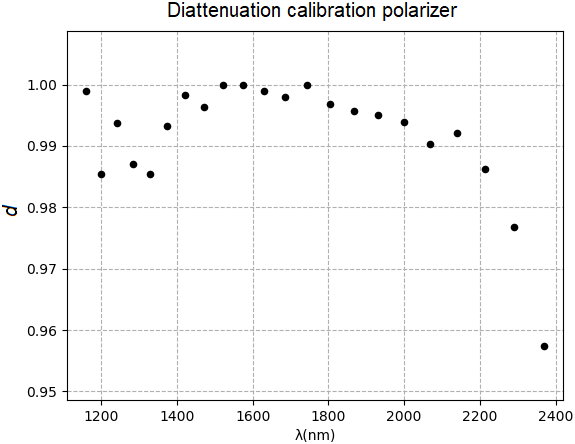}
\end{tabular}
\end{center}
\caption{
Fitted diattenuation of the calibration polarizer as a function of wavelength. Ideally, the diattenuation would have a value of 1 for all wavelengths.} 
\label{fig:MeasuredDiattenuation}
\end{figure}

In \autoref{fig:MeasuredDiattenuation} we show the fitted values for the diattenuation of the calibration polarizer $d$ as a function of wavelength.  
It shows that the diattenuation is relatively constant ($0.98 < d \leq 1$) between 1180 and 2200 nm and drops sharply for $\lambda > 2200$ nm. 
This sharp drop can be explained by a sudden decrease in optical performance or, more likely, by a measurement bias due to the increasing thermal background of the instrument at these longer wavelengths. 
For the offset angle of the calibration polarizer, we assumed a constant value with wavelength and find $\delta_\mathrm{cal} = 0.76^\circ$.

\begin{figure}[!htb]
\begin{center}
\begin{tabular}{c}
\includegraphics[width=1.\textwidth]{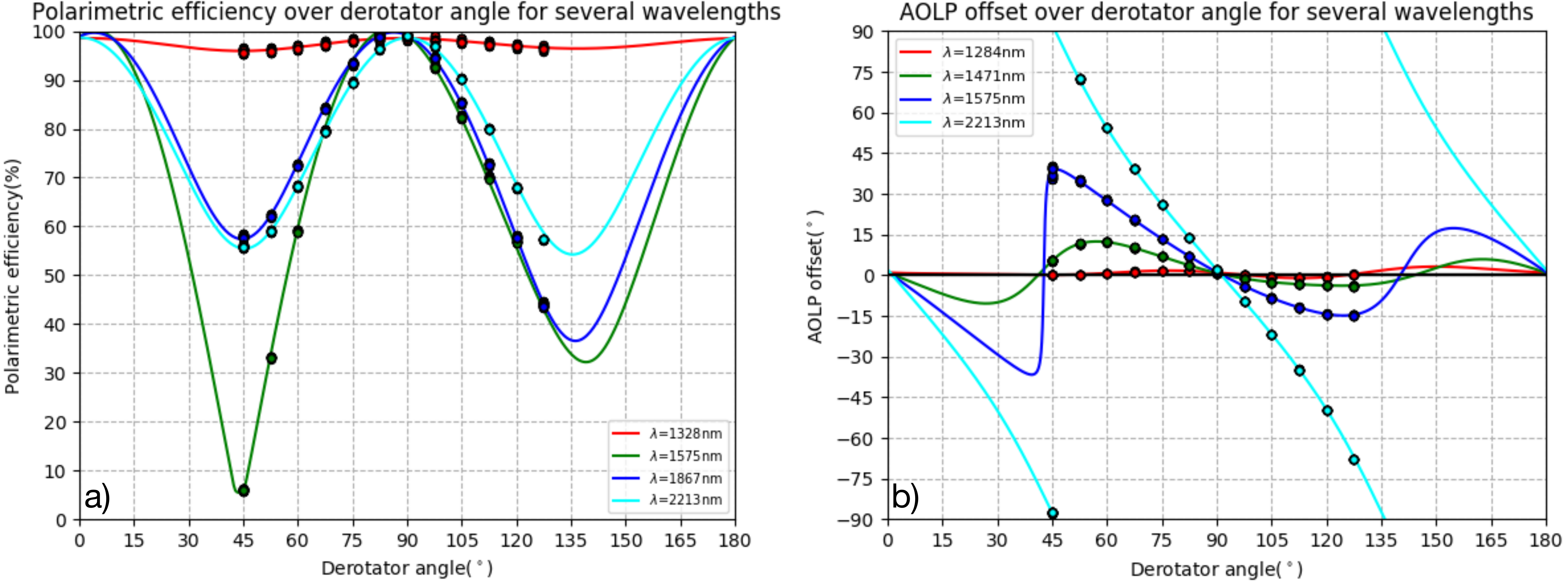}
\end{tabular}
\end{center}
\caption{
Measured a) polarimetric efficiency and b) offset of the angle of linear polarization (AOLP) as a function of derotator angle for various wavelengths. Ideally, the polarimetric efficiency would be 100\% and the offset of the angle of linear polarization $0^\circ$ for all wavelengths and derotator angles. We note that the two panels show slightly different wavelength bins.}
\label{fig:MeasuredPolarimetricEfficiency}
\end{figure}

By comparing the measured degree of linear polarization $P = \surd{(q^2 + u^2)}$ and angle of linear polarization $\chi = {}^1{\mskip -3mu/\mskip -1mu}_2 \arctan(u / q)$ with the values these quantities would have if the optical components were ideal, we can calculate the polarimetric efficiency and offset of the angle of linear polarization.
The results of these calculations for various wavelengths are shown as a function of derotator angle in \autoref{fig:MeasuredPolarimetricEfficiency}.
In the ideal case, the polarimetric efficiency would be 100\% for all derotator angles and wavelengths. 
However, as shown in \autoref{fig:MeasuredPolarimetricEfficiency}a, the polarimetric efficiency shows strong variations with derotator angle.
The polarimetric efficiency is minimized for derotator angles around $45^{\circ}$ and $135^{\circ}$, and is maximized for derotator angles close to $0^{\circ}$, $90^{\circ}$ and $180^{\circ}$. 
This behaviour is due to the derotator having a retardance that is not equal to 180$^{\circ}$ (see \autoref{fig:MeasuredRetardances}a). 
For derotator angles close to 45$^{\circ}$ and 135$^{\circ}$, the derotator is optimally oriented to convert the horizontal and vertical polarization states to circular polarization, which cannot be measured using the Wollaston prism.
The asymmetry between the minima of the polarimetric efficiency at derotator angles of 45$^{\circ}$ and 135$^{\circ}$ can be explained by the HWP retardance being offset from 180$^\circ$.

\begin{figure}[!htb]
\begin{center}
\begin{tabular}{c}
\includegraphics[width=0.5\textwidth]{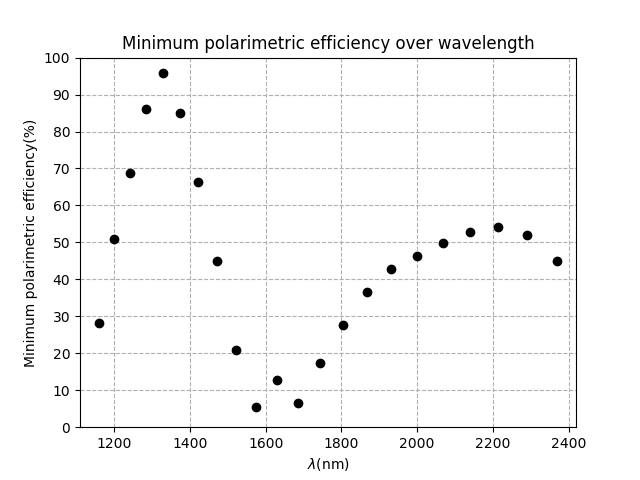}
\end{tabular}
\end{center}
\caption{
Minimum polarimetric efficiency as a function of wavelength.} 
\label{fig:MinEff}
\end{figure}

From \autoref{fig:MeasuredPolarimetricEfficiency}a, we also see that the derotator retardance varies strongly with wavelength.
For $\lambda = 1575$ nm the polarimetric efficiency even drops to a value close to 5\% at a derotator angle of 45$^{\circ}$.
An overview of the minimum polarimetric efficiency as a function of wavelength is plotted in \autoref{fig:MinEff}. 
This figure shows that the polarimetric efficiency peaks around $\lambda = 1300$ nm, when the derotator retardance is closest to 180$^{\circ}$ (see \autoref{fig:MeasuredRetardances}a).
We also see that at wavelengths around $\lambda=1600$ nm the lowest polarimetric efficiency can be reached, because at these wavelengths the derotator acts as an almost perfect QWP.

For an ideal instrument, there would be no offset of the angle of linear polarization for all derotator angles and wavelengths. 
However, as shown in \autoref{fig:MeasuredPolarimetricEfficiency}b, the derotator retardance does not only decrease the polarimetric efficiency, but also introduces a non-zero offset of the angle of linear polarization. 
The offset is strongest for derotator angles close to 45$^{\circ}$ and 135$^{\circ}$ where the polarimetric efficiency is lowest, and is close to zero for derotator angles around 0$^{\circ}$, 90$^{\circ}$, and 180$^{\circ}$.
Similar to the curves of the polarimetric efficiency, the curves for the angle offset show an asymmetry between derotator angles of 45$^{\circ}$ and 135$^{\circ}$ that is caused by the non-ideal value of the HWP retardance.
As with the polarimetric efficiency, we find that the angle offset strongly varies with wavelength. 
For several wavelength bins the offset angle can even be as large as $\pm90^{\circ}$.

Finally, we note that the results discussed in this subsection are very similar to those presented for SPHERE-IRDIS in Ref~\citenum{van2020polarimetric}.
One reason for this is that the optical setups of SPHERE-IRDIS and SCExAO-CHARIS are quite similar.
However, the results also suggests that the design of the HWP and the coating of the derotator are comparable to those used for SPHERE. 

\subsection{Instrumental polarization of the telescope}

The instrumental polarization (IP) of the telescope is almost completely created by the telescope's tertiary mirror (M3). 
This silver-coated mirror has an angle of incidence of $45^\circ$ and deflects the light arriving from the secondary mirror to the Nasmyth platform where SCExAO is located. 
The first and secondary mirror of the telescope are axisymmetric and therefore do not create significant instrumental polarization\cite{tinbergen2005astronomical}.

We can theoretically predict the IP of the telescope by computing the diattenuation $\epsilon$ of M3 for each wavelength bin with the Fresnel equations.
For this calculation we obtain the refractive indices of silver from Ref.~\citenum{rakic1998optical}. 
We describe the measurements with only the part of \autoref{eq:complete_model} upstream of the HWP:
\begin{align}
	\boldsymbol{S}_\mathrm{HWP} = T(-a) M_\mathrm{tel} T(p) \boldsymbol{S}_\mathrm{in}.
\label{eq:telescope_model} 
\end{align} 
We set $\boldsymbol{S}_\mathrm{in} = [1, 0, 0, 0]^T$, and retrieve the normalized Stokes parameters $q$ and $u$ from the second and third elements of $\boldsymbol{S}_\mathrm{HWP}$.

\begin{figure}[!htb]
\begin{center}
\begin{tabular}{c}
\includegraphics[width=\textwidth]{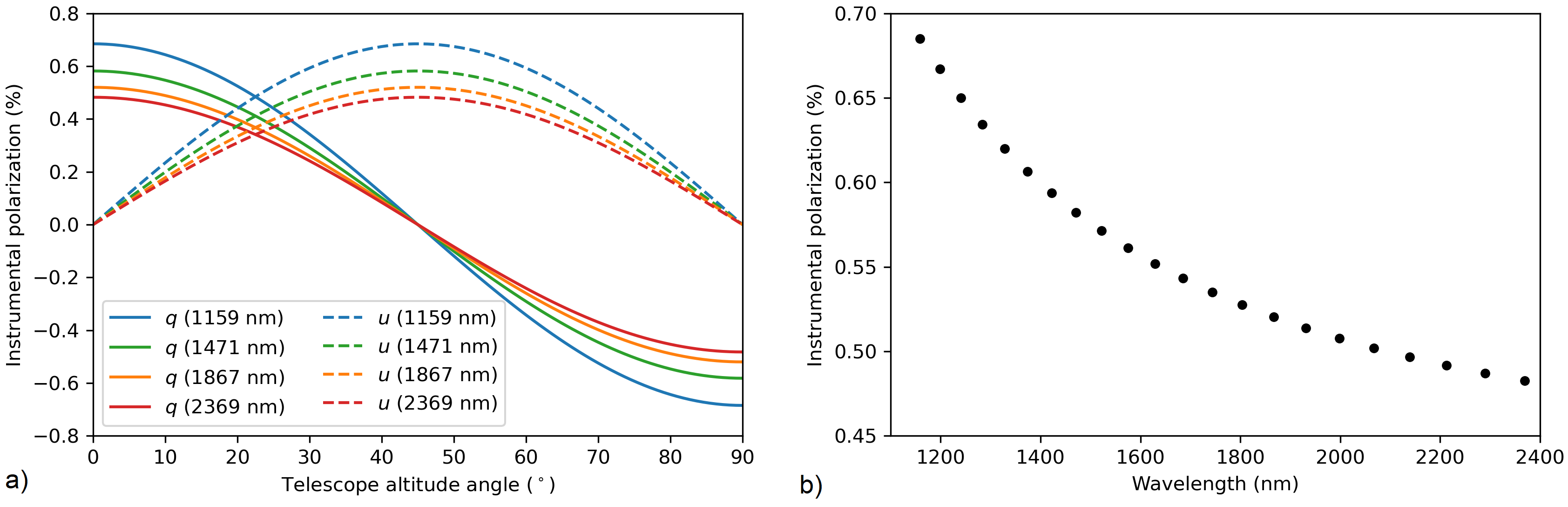}
\end{tabular}
\end{center}
\caption{Theoretically predicted instrumental polarization of the telescope expressed as a) the normalized Stokes parameters $q$ and $u$ as a function of telescope altitude angle for four wavelength bins, and b) the diattenuation for the 22 wavelength bins.} 
\label{fig:ip_telescope}
\end{figure}

\autoref{fig:ip_telescope}a shows the resulting $q$ and $u$ as a function of telescope altitude angle for four of the wavelength bins, including the bins of the shortest and longest wavelengths.
The shape of the curves can be explained as follows.
M3 produces IP that is oriented perpendicular to the plane of incidence of the mirror, and this plane of incidence rotates with respect to the instrument when the telescope altitude angle is changed.
Whereas the degree of linear polarization $P = \epsilon$ of this IP signal does not change with altitude angle, the angle of linear polarization is exactly equal to the altitude angle: $\chi = a$.
The normalized Stokes parameters thus vary as $q = \epsilon\cos(2a)$ and $u = \epsilon\sin(2a)$.
We note that the shape of the curves for $q$ and $u$ differ from those determined for SPHERE-IRDIS in Ref~\citenum{van2020polarimetric}, because in that case the IP of both the telescope and the first mirror of SPHERE (M4) are considered.

From \autoref{fig:ip_telescope}a we also see that the IP is different for the various wavelength bins plotted.
In \autoref{fig:ip_telescope}b we therefore show the IP (the diattenuation) for each of the 22 wavelength bins.
The IP is largest for the shortest wavelength (${\sim}0.7\%)$, and decreases monotonically for longer wavelengths to a value of ${\sim}0.5\%$.
This behavior is exactly the same as that found for the telescope of SPHERE-IRDIS in Ref~\citenum{van2020polarimetric}.
However, in that case the IP is overall larger, because the M3 of the Very Large Telescope is coated with aluminum rather than silver.

Although the results presented in \autoref{fig:ip_telescope} are qualitatively accurate, the exact values for the diattenuation at each wavelength bin will differ in reality.
We have therefore observed the polarization standard star HD~283809\cite{messinger1997interstellar} during the SCExAO engineering nights on 31 January, and 9 February 2020.
This polarization standard star has a well-measured degree and angle of linear polarization that we will use to accurately calibrate the IP of the telescope as a function of wavelength.

%
%

\section{Conclusions and outlook}
\label{sec:DiscussionOutlook}

In this manuscript we have presented the first results of our efforts to characterize the instrumental polarization effects of the spectropolarimetric mode of SCExAO-CHARIS. 
We have created a detailed Mueller matrix model describing the telescope and instrument.
Using measurements with the internal light source and a calibration polarizer, we have determined the retardance of the derotator and the HWP for each of the 22 wavelength bins.
We find that the retardance of the derotator strongly varies with wavelength (with values between 50$^\circ$ and 250$^{\circ}$), and that it acts as an almost perfect quarter-wave plate (QWP) at wavelengths around 1600 nm. 
The retardance of the HWP varies between 167.5$^{\circ}$ and 185$^{\circ}$ over the wavelength range considered.
The curve of the retardance versus wavelength suggests that the HWP is an achromatic wave plate, similar to those commercially available. 
The non-ideal values of the derotator retardance results for some wavelengths in very low polarimetric efficiencies at derotator angles around 45$^\circ$ and 135$^\circ$.
In the worst case, at wavelengths close to 1600 nm, $\sim$95\% of the incident linear polarization is converted to circular polarization that cannot be measured. 
The variations in the polarimetric efficiency are accompanied by offsets of the angle of linear polarization that can be as large as $\pm90^\circ$.
These results are very similar to those obtained with calibrations of SPHERe-IRDIS as presented in Ref~\citenum{van2020polarimetric}.

To estimate the instrumental polarization (IP) of the telescope, we have performed theoretical calculations for each wavelength bin using the Fresnel equations.
We find that the degree of linear polarization of the IP does not vary with the telescope altitude angle. This degree of linear polarization is largest (${\sim}0.7\%$) for the shortest wavelengths, and decreases monotonically for longer wavelengths to a value of ${\sim}0.5\%$. 
The angle of linear polarization of the IP is exactly equal to the altitude angle of the telescope.

In the future, we plan to accurately determine the IP of the telescope with measurements of the polarization standard star HD~283809 that we recently obtained.
We will also fit comprehensive physical models to all calibration data, including models that accurately describe the design of the HWP and the coating of the derotator. 
Subsequently, we will perform a careful error analysis to determine the polarimetric accuracy of our model.
Finally, we will integrate the complete Mueller matrix model into the existing CHARIS post-processing pipeline\cite{currie2012direct, currie2011, currie2018scexao}, with the aim to achieve a polarimetric accuracy of ${<}0.1\%$ in the degree of linear polarization.
This pipeline will be comparable to the IRDAP\cite{van2020polarimetric}\footnote{\url{https://irdap.readthedocs.io/}} pipeline for SPHERE-IRDIS, and will be made publicly available to the community.  

The spectropolarimetric mode of CHARIS is one of three polarimetric modes currently available at SCExAO.
A fast polarimetry mode combined with non-redundant aperture masking interferometry is offered in the visible (600-800 nm) by the VAMPIRES instrument\cite{norris2015vampires}. 
Currently under development is a fast NIR polarimetry mode employing a ferroelectric liquid crystal that provides fast polarization modulation (up to a $\sim$1 kHz) in synchronisation with a C-RED ONE camera\cite{feautrier2017c}, and a Wollaston prism to split the orthogonal linear polarization states.
Compared to CHARIS, these two additional modes operate with different polarization-sensitive components, and use either different wavelength ranges or different bandwidths. 
To enable accurate polarimetry with these modes as well, separate polarimetric calibration measurements need to be performed. 
Our calibrations of CHARIS' spectropolarimetric mode will enable unique quantitative polarimetric studies of circumstellar disks and planetary and brown dwarf companions. 

\acknowledgments 

R.G. van Holstein thanks ESO for the studentship at ESO Santiago during
which part of this project was performed. The research of S.P. Bos and F. Snik leading to these results has received funding from the European Research Council under ERC Starting Grant agreement 678194 (FALCONER). 
The development of SCExAO was supported by the Japan Society for the Promotion of Science (Grant-in-Aid for Research \#23340051, \#26220704, \#23103002, \#19H00703 \& \#19H00695), the Astrobiology Center of the National Institutes of Natural Sciences, Japan, the Mt Cuba Foundation and the director's contingency fund at Subaru Telescope.
The authors wish to recognize and acknowledge the very significant cultural role and reverence that the summit of Maunakea has always had within the indigenous Hawaiian community. 
We are very fortunate to have the opportunity to conduct observations from this mountain. 

\bibliography{main} 
\bibliographystyle{spiebib} 

\end{document}